\documentstyle[12pt,epsf]{article}
\baselineskip=\normalbaselineskip
\ifx\TwoupWrites\UnDeFiNeD\else\target{\magstepminus1}{11.3in}{8.27in}
	\source{\magstep0}{7.5in}{11.69in}\fi
\newfont{\fourteencp}{cmcsc10 scaled\magstep2}
\newfont{\titlefont}{cmbx10 scaled\magstep3}
\newfont{\authorfont}{cmcsc10 scaled\magstep1}
\newfont{\fourteenmib}{cmmib10 scaled\magstep2}
	\skewchar\fourteenmib='177
\newfont{\elevenmib}{cmmib10 scaled\magstephalf}
	\skewchar\elevenmib='177
\makeatletter
\newcommand\nonsequentialeqnum{
	\@addtoreset{equation}{section}
	\def\theequation{\arabic{section}.\arabic{equation}}}
\newif\ifp@bblock  \p@bblocktrue
\newcommand\nopubblock{\p@bblockfalse}
\newcommand\topspace{\hrule height 0pt depth 0pt \vskip}
\newcommand\p@bblock{\begingroup \tabskip=\hsize minus \hsize
	\baselineskip=1.5\ht\strutbox \topspace-2\baselineskip
	\halign to\hsize{\strut ##\hfil\tabskip=0pt\crcr
	\the\Pubnum\crcr\the\date\crcr}\endgroup}

\renewcommand\titlepage{\ifx\TwoupWrites\UnDeFiNeD\null\vspace{-1.7cm}\fi
\vskip0.6cm
	\ifp@bblock\p@bblock \else\hrule height 0pt \relax \fi}
\makeatother
\newtoks\date
\newtoks\Pubnum
\newtoks\pubnum
\Pubnum={
YITP/K-1045 \crcr
KEK-TH-381 \crcr
KEK-Preprint 93-172 \crcr
YITP/U-\the\pubnum\crcr
hep-th/9312175
}
\date={\today}
\newcommand{\frontpageskip}{\vspace{12pt plus .5fil minus 2pt}}
\renewcommand{\title}[1]{\frontpageskip
	\begin{center}{\titlefont #1}\end{center}\par}
\renewcommand{\author}[1]{\frontpageskip\par\begin{center}
	{\authorfont #1}\end{center}
	\nobreak
	}

\newcommand{\address}[1]{\par\begin{center}{\sl #1}\end{center}\par}

\renewcommand{\thanks}[1]{\footnote{#1}}
\renewcommand{\abstract}{\par\frontpageskip\centerline{\fourteencp Abstract}
	\vspace{8pt plus 3pt minus 3pt}}

\begin{document}
%
\pubnum{93-37}
\date{December 1993 \crcr (Revised)}
\titlepage

\renewcommand{\thefootnote}{\fnsymbol{footnote}}
\title{
Two-Dimensional Quantum Gravity\\
\vskip5pt
in Temporal Gauge
}

\author{
M.\ Fukuma${}^{1\,}$\thanks{
e-mail address: fukuma@yukawa.kyoto-u.ac.jp},
N.\ Ishibashi${}^{2\,}$\thanks{
e-mail address: ishibash@theory.kek.jp},
H.\ Kawai${}^{2\,}$\thanks{
e-mail address: kawaih@theory.kek.jp}
and M.\ Ninomiya${}^{3\,}$\thanks{
e-mail address: ninomiya@yukawa.kyoto-u.ac.jp}
}

\address{
${}^1$
Yukawa Institute for Theoretical Physics\\
Kyoto University, Kyoto 606, Japan \\
${}^2$
National Laboratory for High Energy Physics (KEK)\\
Tsukuba, Ibaraki 305, Japan\\
${}^3$
Uji Research Center, Yukawa Institute for Theoretical Physics\\
Kyoto University, Uji 611, Japan
}

\renewcommand{\thefootnote}{\arabic{footnote}}
\setcounter{footnote}{0}
\newcommand{\cleqn}{\setcounter{equation}{0} \indent}
\renewcommand{\theequation}{\thesection.\arabic{equation}}
\newcommand{\beqa}{\begin{eqnarray}}
\newcommand{\eeqa}{\end{eqnarray}}
\newcommand{\n}{\nonumber}
\newcommand{\nn}{\nonumber \\ }
\newcommand{\eq}[1]{(\ref{#1})}
\newcommand{\cD}{{\cal D}}
\newcommand{\cH}{{\cal H}}
\newcommand{\Psid}{\Psi^\dagger}
\newcommand{\norm}[1]{{\parallel {#1} \parallel}^2}
\newcommand{\nnorm}[1]{{{\parallel {#1} \parallel}^{\prime\,2}_l}}
\newcommand{\del}{\partial}
\newcommand{\db}{{\bar{\delta}}}
\newcommand{\gbar}{{\bar{g}}}
\newcommand{\dl}
           {\left[\,\frac{dl}{\,l\,}\,\right]}
\newcommand{\Det}{\,\mbox{Det}\,}
\newcommand{\Tr}{\,\mbox{Tr}\,}
\newcommand{\ldot}{\dot{l}}
\newcommand{\const}{\mbox{const.\ }}
\newcommand{\bra}[1]{\left\langle\,{#1}\,\right|}
\newcommand{\ket}[1]{\left|\,{#1}\,\right\rangle}

\begin{abstract}

We propose a new type of gauge in two-dimensional quantum gravity.
We investigate pure gravity in this gauge, and find that the system 
reduces to quantum mechanics of loop length $l$.
Furthermore, we rederive the $c\!=\!0$ string field theory which was
discovered recently.
In particular, the pregeometric form of the Hamiltonian is naturally
reproduced.
~\\
~\\
\begin{flushleft}
{\sc To be published in} {\it Nuclear Physics} {\bf B}
\end{flushleft}
\end{abstract}

\newpage

\section{Introduction}
\cleqn

Two-dimensional (2D) quantum gravity has been extensively studied
for the last few years, since it serves as a toy model of 4D quantum
gravity as well as a prototype of string theories \cite{p}.
Recent developments in 2D quantum gravity are mainly due to the
discovery of the dynamical triangulation method \cite{rtm}.
In particular, the success of summing over all topologies in matrix
models \cite{bk}\cite{ds}\cite{gm} have attracted much interest in this
area.

The major advantage of the matrix model approach is that one can
calculate all the correlation functions for any genus, and then
sum them up in such a way that the factorization property holds.
Furthermore, the Schwinger-Dyson equation, which contains all the
information in the matrix models, revealed some universal structure
of 2D quantum gravity \cite{fkn}\cite{dvv}.

In spite of such great success, there remain some difficulties in
the dynamical triangulation method:
Firstly, it is hard to identify the operators constructed in the
lattice approach with those in the continuum theory.
Secondly, there appears a strong restriction on matters coupled to
gravity.
In fact, until now we have no way to couple the matters whose central
charge is greater than one.
Thirdly, although the dynamical triangulation method can be extended to
higher dimensions \cite{a}, analytic calculation has been
successfully performed only in two dimensions.
All these facts naturally lead us to investigate 2D quantum gravity
in continuum approach.

There have been mainly two continuum approaches so far.
One is based on the conformal gauge \cite{p}\cite{ddk}, and the other
on the light-cone gauge \cite{kpz} (see, {\em e.g.}, ref.\ \cite{dp}
for further references).
Although an ADM (Arnowitt-Deser-Misner)-like formalism may be
useful for the physical interpretation of the dynamics,
these two gauges are not suitable for this formalism.
Moreover, they do not work in higher dimensions except for the
vicinity of two dimensions \cite{kkn}.

Recently a new approach is advocated for 2D quantum gravity by two of
the present authors \cite{ik}.
In that work 2D quantum gravity was formulated so that
minisuperspace approximation is exact, {\em i.e.}, the system 
reduces to quantum mechanics of loop length $l$, as was demonstrated
in the transfer-matrix formalism of 2D quantum gravity which was
initiated in ref.\ \cite{kkmw}.\footnote{
Minisuperspace approximation is originally taken in quantum Liouville
theory to investigate loop dynamics \cite{mss}.
}

Let us review here the major consequence of ref.\ \cite{ik}.
First one introduces loop-annihilation and -creation operators,
$\Psi(l)$ and  $\Psid(l)$, which satisfy the commutation relation
\beqa
   ~\left[\,\Psi(l),\,\Psid(l')\,\right]~=~\delta(l-l')\,. \label{1.1}
\eeqa
Then the following form of Hamiltonian is assumed in the second
quantization formalism:
\beqa
   \cH&=&\int dl\,dl'\,K(l,l')\,\Psid(l)\,\Psi(l') \nn
   &&~~+\,\int dl\,dl'\,(l+l')\,\Psid(l)\,\Psid(l')\,\Psi(l+l') \nn
   &&~~+\,g\,\int dl\,dl'\,l\,l'\,\Psid(l+l')\,\Psi(l)\,\Psi(l') \\
   &&~~+\,\int dl \rho(l)\,\Psi(l)\,. \n
\eeqa
The first term (kinetic term) represents the amplitude of
cylinder where initial and final loop lengths are $l'$ and $l$.
The second term describes the splitting of loop, and the third term
describes the merging of loops, so that $g$ is identified with
the renormalized string coupling constant.
The fourth term represents cap amplitude, {\it i.e.}, the
amplitude for instantaneous vanishing of loop with length $l$.
The main result in ref.\ \cite{ik} is that the Schwinger-Dyson
equation for $c=0$ matrix model is completely reproduced if we assume
$K(l,l')\equiv0$ (pregeometric type) and
$\rho(l)\equiv\delta''(l)-\mu\delta(l)$ with $\mu$ the renormalized
cosmological constant.

The above string field Hamiltonian was derived via the dynamical
triangulation approach \cite{ik}. 
In the present paper, we will attempt to rederive it from the
continuum approach.\footnote{
Recently, the string field Hamiltonian was also rederived from the
collective field theory approach to the matrix models \cite{jr}. 
The fictitious time they introduced for the stochastic quantization of
the matrix models coincides with the time coordinate considered in
ref.\ \cite{ik} using the geodesic distance. 
}
We propose a new type of gauge (``temporal'' gauge) and show that the
above Hamiltonian $\cH$ is naturally obtained in this gauge.
In particular, the vanishing of the kinetic term, $K(l,l')=0$, will be
demonstrated.
In the following, pure gravity is investigated, while
inclusion of matters will be reported elsewhere.

The present paper is organized as follows:
In sect.\ 2, we introduce our temporal gauge by using the ADM
decomposition.
In sect.\ 3, we briefly describe the calculation of the cylinder
amplitude, and then show that 2D pure gravity actually
reduces to quantum mechanics of loop length $l$.
In sect.\ 4, we discuss how the second-quantized Hamiltonian $\cH$ is
obtained in our formulation.
In particular, we present the mechanism for vanishing of the kinetic term.
Sect.\ 5 is devoted to conclusions.

\section{ADM decomposition and temporal gauge}
\cleqn

It is known that the ADM decomposition is useful for the Hamiltonian
formalism of quantum gravity, for which a metric $g_{\mu\nu}$ (with
Euclidian signature) on a two-dimensional manifold is parametrized as
\beqa
	ds^2&=&g_{\mu\nu}\,dx^\mu dx^\nu\nn
	    &=&(Ndx^0)^2\,+\,h\,(\lambda dx^0+dx^1)^2, \label{2.1}
\eeqa
{\it i.e.\,},
\beqa
 	g_{\mu\nu}~=~
        \left[\begin{array}{cc}
	N^2+h\lambda^2 & h\lambda \\
        h\lambda & h
	\end{array}\right]\, . \label{2.2}
\eeqa
Here $N(x^0,x^1)$ is the lapse function, $\lambda(x^0,x^1)$ the shift
function and $h(x^0,x^1)$ the metric on the time slice at $x^0$.
Eq.\ \eq{2.1} is nothing but Pythagoras' theorem;
two time slices at $x^0$ and $x^0+dx^0$, respectively, are separated
from each other by geodesic distance $Ndx^0$, while the space coordinate
$x^1$ shifts by $\lambda dx^0$ under such time evolution.

In this section as well as the following one, we consider a cylinder
$M$ with two loop boundaries $C$ and $C'$.
The cylinder can be regarded as the world sheet swept by a loop under
its time evolution, and we call $C$ the incoming loop and $C'$ the
outgoing loop.
We further assume that there occurs no splitting of loop, and those
two loops $C$ and $C'$ are separated from each other by geodesic
distance $D$ (see Fig.\ 1).

To state our assumption more precisely, we first prepare some
notations.
Let $d\,(p,q)$ be the geodesic distance between two points $p,\,q\,\in
M$.
Let then $d\,(p\,;S)$ be the minimal geodesic distance between a point
$p\in M$ and a subset $S\subset M$,
\beqa
   d\,(p\,;S)~\equiv~\inf_{q\in S}\,d\,(p,q)\,. \label{2.a}
\eeqa
Furthermore, we denote by $C_d$ the subset consisting of the points
that are separated from the incoming loop $C$ by geodesic distance
$d$:
\beqa
   C_d~\equiv~\left\{\,p\in M\,|\,d\,(p\,;C)=d\,\right\}\,. \label{2.b}
\eeqa
Note here that $C_0 = C$, and also that each subset $C_d$ can always
be identified with some time slice if we apply the ADM decomposition of
metric to a coordinate system in which $C$ is represented by
$x^0=\const$.
Thus, the assumption we made above can be rephrased as that the
metrics $g_{\mu\nu}$ on $M$ satisfy the following two conditions:
\beqa
   \mbox{(\romannumeral1)} && \mbox{$C_d$ is homeomorphic to $C$
   ~~~$(0\leq {}^\forall d\leq D)$}\,, \nn
   \mbox{(\romannumeral2)} && C_D~=~C'\,. \label{2.c}
\eeqa

Now we are going to calculate the following functional integral over
those metrics that satisfy the condition \eq{2.c}:
\beqa
   f(l',l;D)&\equiv&\int
   \frac{\cD g_{\mu\nu}}{\mbox{Vol(diff)}}\,
   \exp\left\{-\mu_0\int d^2x\sqrt{g}\right\} \nn
   &&~~\cdot\,\delta\left(\int_C\sqrt{g_{\mu\nu}dx^\mu
   dx^\nu}-l\right)\,
   \delta\left(\int_{C'}\sqrt{g_{\mu\nu}dx^\mu dx^\nu}-l'\right)
   \label{2.3} \\
   &&~~\cdot\,\theta\,\left[\,\mbox{condition \eq{2.c}}\,\right]\,, \n
\eeqa
where $\mu_0$ is the bare cosmological constant.
In this expression, the functional measure $\cD g_{\mu\nu}$ is defined
via the positive definite norm of the infinitesimal deformation of
metric around $g_{\mu\nu}$ as follows:
\beqa
   \norm{\delta g_{\mu\nu}}_g~\equiv~\int d^2x\,\sqrt{g}\,
   g^{\mu\alpha}g^{\nu\beta}\,\delta g_{\mu\nu}|_g\,\delta
   g_{\alpha\beta}|_g\,.  \label{2.4}
\eeqa
The volume of diffeomorphism group Vol(diff) is identified with
the functional integral over all diffeomorphisms,
\beqa
   \mbox{Vol(diff)}~\equiv~\int\cD f^\mu\,, \label{2.5}
\eeqa
where $\cD f^\mu$ is determined by the norm of infinitesimal
diffeomorphism $(x^\mu\,\mapsto\,x^\mu-\delta f^\mu(x))$ with
\beqa
   \norm{\delta f^\mu }_g~=~\int d^2x\,\sqrt{g}\,g_{\mu\nu}\,
   \delta f^\mu \delta f^\nu. \label{2.6}
\eeqa

We parametrize the gauge slice by $\gbar_{\mu\nu}$, and
impose on it the following ``temporal'' gauge condition:
\beqa
   \bar{N}&\equiv&1\,, \label{2.7} \\
   \del_1\bar{h}&\equiv&0\,, \label{2.8}
\eeqa
or equivalently,
\beqa
   \gbar_{\mu\nu}(x^0=t,x^1=x)~\equiv~
   \left[\begin{array}{cc}
    1+l(t)^2k(t,x)^2 & l(t)^2k(t,x) \\
    l(t)^2k(t,x) & l(t)^2
   \end{array}\right]. \label{2.9}
\eeqa
Here we have introduced $l(t)^2$ ($=\,\bar{h}$) as an integration
constant for eq.\ \eq{2.8}, and denote
$\bar{\lambda}(t,x)$ by $k(t,x)$ for simplicity.

The geometrical meaning of this condition should be clear:
Eq.\ \eq{2.7} implies that the time coordinate $x^0$ is chosen
directly as the geodesic distance from the incoming loop $C$, and eq.\
\eq{2.8} ensures that the space-like metric $h$ has no fluctuations
along time slices.
All these properties are consistent with the ones in the
transfer-matrix formalism of 2D quantum gravity based on the
dynamical triangulation method \cite{kkmw}.
In the following, we will take a global coordinate system
$(x^0,x^1)=(t,x)$ on $M$, such that $0\leq x\leq1$, and $C$ and $C'$
are represented by $t=0$ and $t=D$, respectively.

For the parametrization given above, it may be useful to introduce two
vectors $\vec{n}$ and $\vec{s}$ which are normal and tangential to
time slices, respectively, so that
\beqa
   n^\mu~\equiv~(1,-k),~~~s^\mu~\equiv~(0,l^{-1}). \label{2.10}
\eeqa
Note that they satisfy the following orthonormal conditions:
\beqa
   \gbar_{\mu\nu}n^\mu n^\nu&=&\gbar_{\mu\nu}s^\mu s^\nu~=~1, \nn
   \gbar_{\mu\nu}n^\mu s^\nu&=&0. \label{2.11}
\eeqa
It is easy to see that eq.\ \eq{2.5} evaluated at
$g_{\mu\nu}=\gbar_{\mu\nu}$ is rewritten in the following form:
\beqa
   \norm{\delta f^\mu}_\gbar~=~\int d^2x\,l\,
   \left[ (\delta v^n)^2+(\delta v^s)^2\,\right]\,, \label{2.12}
\eeqa
where $\delta v^n$ and $\delta v^s$ are infinitesimal diffeomorphisms in the
normal and tangential directions, respectively, {\it i.e.\,},
\beqa
   \delta v^n~\equiv~\gbar_{\mu\nu}\,n^\mu\,\delta f^\nu\,,~~~
   \delta v^s~\equiv~\gbar_{\mu\nu}\,s^\mu\,\delta f^\nu\,,\label{2.13}
\eeqa
and we impose on them the following boundary conditions:
\beqa
   \delta v^n|_{\del M}~=~0~~~~~~~~~~~~~~~~~~~~~~~&&\mbox{(Dirichlet)}\,, \nn
   \delta v^s(x^0,x^1=0)~=~\delta
   v^s(x^0,x^1=1)~~&&\mbox{(periodic)}\,. \label{2.14} 
\eeqa
Furthermore, since the infinitesimal deformation of metric around
$\gbar_{\mu\nu}$ is generally expressed as\footnote{
$\overline{\nabla}_\mu$ is the covariant derivative with respect to
$\gbar_{\mu\nu}$.
}
\beqa
   \delta g_{\mu\nu}~=~\delta\gbar_{\mu\nu}\,+\,
   \overline{\nabla}_\mu\delta f_\nu\,+\,
   \overline{\nabla}_\nu\delta f_\mu, \label{2.15}
\eeqa
we can also rewrite eq.\ \eq{2.4} into the following form after
straightforward calculation as will be proved in Appendix A:
\beqa
   \norm{\delta g_{\mu\nu}}_\gbar 
   &=& \int dt\,\frac{\delta l(t)^2}{l(t)} 
   \,+\,\frac{1}{2}\int d^2x\,l\,(l\,\delta k)^2 \nn
   &&~~~+\,\int d^2x\,l\,\left[\,
   (D_n\,\delta v^n)^2\,+\,(D_s \,\delta v^s)^2\,\right] \label{2.16}\\
   &=& \int dt\,\frac{\delta l(t)^2}{l(t)} 
   \,+\,\frac{1}{2}\int d^2x\,l\,(l\,\delta k)^2 \nn
   &&~~~+\,\int d^2x\,l\,\left[\,
   \delta v^n\,D_n^\dagger D_n\,\delta v^n\,+\,
   \delta v^s\,D_s^\dagger D_s\,\delta v^s\,\right]\,. \n
\eeqa
Here $D_n$ and $D_s$ are the derivatives in the normal and tangential
directions, respectively: 
\beqa
   D_n&\equiv&n^\mu\del_\mu~=~\del_0\,-\,k\,\del_1, \nn
   D_s&\equiv&s^\mu\del_\mu~=~l^{-1}\,\del_1\,, \label{2.17}
\eeqa
and $D_n^\dagger$, $D_s^\dagger$ are their hermitian conjugates under
the diffeomorphism-invariant measure\footnote{
We use the following abbreviation:
\beqa
   \dot{f}~\equiv~\del_0 f,~~~f'~\equiv~\del_1 f. \n
\eeqa
}~
$\int d^2x\,\sqrt{\gbar}\,=\,\int d^2x\,l$:
\beqa
   D_n^\dagger&=&-\,\del_0\,+\,k\,\del_1\,+\,k'\,-\,\frac{\ldot}{l}\,,\nn
   D_s^\dagger&=&-\,l^{-1}\,\del_1\,.
   \label{2.19} 
\eeqa

Due to the boundary condition \eq{2.14}, the operator $D_n^\dagger D_n$
has no zero-modes while those of $D_s^\dagger D_s$ are infinitely
degenerated, each of them specified by time $t$. 
Thus, to simplify the calculation of the determinant of $D_s^\dagger
D_s$, we consider the following eigenvalue problem with $t$ regarded
as a parameter: 
\beqa
   D_s^\dagger D_s (t)\,\psi_a(x;t)~=~\lambda_a(t)\,
   \psi_a(x;t)~~~~(a\geq0)\,. \label{F1}
\eeqa
Here $D_s^\dagger D_s (t)=-\,l(t)^{-2}\,\del_x^2$, and the wave
functions $\psi_a(x;t)$ are normalized as 
\beqa
   l(t)\,\int_0^1 dx\,\psi_a(x;t)\,\psi_b(x;t)~=~\delta_{ab}\,. 
   \label{F2}
\eeqa 
Note that $D_s^\dagger D_s (t)$ is hermitian under the measure
$\int_0^1 dx$, and the zero-mode at $t$ is given by $\psi_0 =
l(t)^{-1/2}$. 

Now we can give a definite meaning to eq.\ \eq{2.16}. 
We first expand $\delta v^s(t,x)$ as 
\beqa
   \delta v^s(t,x)~=~\sum_{a\geq0} \delta c_a(t)\,\psi_a(x;t)\,. 
\eeqa
Then the tangential components of the norms $\norm{\delta
f^\mu}_{\gbar}$ and $\norm{\delta g_{\mu\nu}}_{\gbar}$ are expressed,
respectively, as 
\beqa
   \int dt\,\sum_{a\geq0}\left( \delta c_a(t)\right)^2~~~~
   \mbox{and}~~~~
   \int dt\,\sum_{a\geq1}\,\lambda_a(t)\,\left( \delta c_a(t)\right)^2.
\eeqa
Thus, we obtain 
\beqa
   \cD g_{\mu\nu}&=&\prod_t \frac{dl(t)}{\sqrt{l(t)}}\cdot
       \cD_lk\cdot\cD v^n\cdot\prod_t\prod_{a\geq1}dc_a(t) \nn
   &&~~~~~~\cdot \Det^{1/2}\,D_n^\dagger D_n\cdot
            \prod_t\Det^{\prime\,1/2}\,D_s^\dagger D_s(t), \label{F3} \\
   \cD f^\mu&=& \cD v^n\cdot\prod_t\prod_{a\geq1}dc_a(t)
   \cdot\prod_t dc_0(t). \label{F4}
\eeqa
A simple calculation shows that 
\beqa
   \prod_t\Det^{\prime\,1/2}\,D_s^\dagger
   D_s(t)&\equiv&\prod_t\prod_{a\geq1}\lambda_a(t)^{1/2} \nn
   &=&\prod_t l(t)\,e^{-\mu_1\int dt\,l(t)}\,, \label{F5}
\eeqa
where $\mu_1$ is a constant depending on ultra-violet cutoff. 
Note that $\int dt\,l(t)$ is nothing but the cosmological term $\int
d^2x \sqrt{\gbar}$ in our temporal gauge. 

The integration over zero-modes, $\prod_t \int dc_0(t)$, reflects the
existence of residual gauge symmetry; 
even after the gauge fixing \eq{2.7} and
\eq{2.8}, we can still twist a loop in the tangential direction at
each time: 
\beqa
   t&\mapsto&t \nn
   x&\mapsto&x-\alpha(t)~~~~(0\leq\alpha(t)<1)\,. 
\eeqa
Since this transformation may be written as $\delta
f^\mu_{\mbox{\scriptsize{res}}}\,=\,\delta^\mu_1\cdot\delta\alpha(t)$,
it will lead to 
\beqa
   \delta v^n_{\mbox{\scriptsize{res}}}&\equiv&n_\mu\,\delta
   f^\mu_{\mbox{\scriptsize{res}}}~=~0\,,\nn
   \delta v^s_{\mbox{\scriptsize{res}}}&\equiv&s_\mu\,\delta
   f^\mu_{\mbox{\scriptsize{res}}}~=~l(t)\,\delta\alpha(t)\,,
\eeqa
which implies that the residual gauge transformation corresponds to
shifting $v^s(t,x)$ by a constant at each time $t$. 
In order to fix such translational symmetry, it is enough to specify
the value of $v^s$ for some spatial coordinate (say, $x_0$) at each
time $t$. 
This is achieved by inserting the following expression into eq.\
\eq{F4}: 
\beqa
   1~=~\prod_t\,\left[\,\int_0^1 d\alpha(t)\,l(t)\,\delta
   \left( v^s(t,x_0)\,-\,l(t)\alpha(t) \right)\,\right]\,. 
   \label{F5.5}
\eeqa
Since $\psi_0=l(t)^{-1/2}$, the delta function can be rewritten as 
\beqa
   \delta\left( v^s(t,x_0)\,-\,l(t)\alpha(t) \right)
   &=&\delta\left( \frac{c_0(t)}{\sqrt{l(t)}}\,+\,\sum_{a\geq1}c_a(t)
   \psi_a(x_0;t)\,-\,l(t)\alpha(t)\right) \nn
   &=&\sqrt{l(t)}\,\delta(c_0(t)\,+\,\cdots )\,.\label{F6} 
\eeqa
Substituting eqs.\ \eq{F5.5} and \eq{F6} into eq.\ \eq{F4}, we obtain 
\beqa
   \cD f^\mu~=~\cD v^n\cdot\prod_t\prod_{a\geq1}dc_a(t)\cdot 
   \prod_t l(t)^{-3/2}\,. \label{F7}
\eeqa
Here we have omitted the ($l$-independent) factor $\prod_t\int_0^1
d\alpha(t)$. 

Combining eqs.\ \eq{F3}, \eq{F4}, \eq{F5} and \eq{F7}, we thus obtain 
\beqa
   \frac{\cD g_{\mu\nu}}{\mbox{Vol(diff)}}&\equiv&
   \frac{\cD g_{\mu\nu}}{\cD f^\mu} \nn
   &=&\dl\cdot\cD_l k\cdot\Det^{1/2}D_n^\dagger
   D_n \cdot e^{-\mu_1\,\int dt\,l(t)}\,. \label{F8}
\eeqa
Here we have introduced the following symbol:
\beqa
   \dl~\equiv~\prod_t\,\frac{dl(t)}{l(t)}\,. \label{2.20}
\eeqa
Recall that the measure $\cD_l k$ is defined by the norm 
\beqa
   \norm{\delta k}_l~=~\int d^2x\,l\,\left(l\,\delta k\right)^2\,. 
\eeqa

\section{Integration over shift function and the cylinder amplitude}
\cleqn

What remains now is to evaluate the following functional of $l(t)$:
\beqa
   F[\,l\,]~\equiv~\int\cD_lk\,\Det^{1/2}\,D_n^\dagger D_n[l,k]\,. 
   \label{E1}
\eeqa 
To do so, we first introduce a new measure $\cD_l'k$ for the shift
function $k$ which is defined by the following norm: 
\beqa
   \nnorm{\delta k}~\equiv~\int d^2x\,l\,\left(\delta k\right)^2\,.
   \label{E2}
\eeqa
This measure is special in the sense that it is invariant under the
following (infinitesimal) transformation:
\beqa
   \db l(t)&\equiv&l(t)^2\,\del_t\,\left(\frac{\db\rho(t)}{l(t)}\right)
   \,,\nn
   \db k(t,x)&\equiv&-\,\del_t\,\left(k(t,x)\,\db\rho(t)\right)\,,
   \label{E3}
\eeqa
where $\db\rho(t)$ is an arbitrary ($t$-dependent) function satisfying
the condition $\db\rho(0) = \db\rho(D) = 0$.\footnote{
In the following, we set $0\leq t\leq D,~0\leq x\leq1$. 
}
In fact, one can easily show that 
\beqa
   \db\,\nnorm{\delta k}~=~
   -\,\int d^2x\,\del_t\,\left[\,l(t)\,\db\rho(t)\,
   \left(\delta k(t,x) \right)^2\,\right]~=~0\,.
\eeqa 
This $\db$-transformation is attributed to a Weyl rescaling of the
metric, $\gbar_{\mu\nu} \mapsto \exp(2\db\dot{\rho}(t)) \cdot
\gbar_{\mu\nu}$, accompanied with a diffeomorphism $(t,x) \mapsto
(t+\db\rho(t), x)$ such that our temporal-gauge condition is restored.
A detailed explanation will be given in Appendix B. 

Since the transformation of measure from $\cD_lk$ to $\cD_l' k$ may be
regarded as a Weyl transformation, the Jacobian is expressed by the
Liouville action  
\beqa
   S_L [\phi;\,g_{\mu\nu} ]\, =\, \int d^2x\, \sqrt{g}\, \left\{
   \frac{1}{2}g^{\mu\nu} \del_\mu \phi \del_\nu \phi\, -\, R_g \phi
   \right\}  
\eeqa
as follows \cite{mmdk}:
\beqa
   \cD_l k&=&\cD_l'k\,\exp\left\{-\frac{1}{48\pi}S_L[\phi=\ln l;
   \,\gbar_{\mu\nu}]\right\} \nn
   &=&\cD_l'k\,\exp\left\{\frac{1}{32\pi}\int_0^D\,dt\,
   \frac{\ldot^2}{l}\right\}\,. 
\eeqa
Thus, $F[\,l\,]$ is expressed as 
\beqa
   F[\,l\,]~=~\exp\left\{\frac{1}{32\pi}\int_0^D\,dt\,
   \frac{\ldot^2}{l}\right\}\,
   \exp\left\{-W[\,l\,]\right\}\,,\label{E4}
\eeqa
where we have introduced 
\beqa
   \exp\{-W[\,l\,]\}~\equiv~\int \cD_l'k\,\Det^{1/2} D_n^\dagger D_n[l,k]\,.
   \label{E5}
\eeqa

The differential operator $D_n^\dagger D_n$ is not elliptic, and thus
need a special care in defining its determinant. 
We here notice that the Laplacian 
\beqa
   \Delta[l,k]&\equiv&-\frac{1}{\sqrt{\gbar}}\del_\mu
   \left(\sqrt{\gbar}\gbar^{\mu\nu}\del_\nu\right) \nn
   &=&D_n^\dagger D_n\,+\,D_s^\dagger D_s \label{3.1}
\eeqa
satisfies the following equation for any positive constant $\beta$:
\beqa
   \Delta[\beta^{-1}l,k]~=~D_n^\dagger D_n\,+\,
   \beta^2D_s^\dagger D_s\,.
\eeqa
Using this equation, we thus can define $D_n^\dagger D_n$
as the limit of the elliptic operator $\Delta$:
\beqa
   D_n^\dagger D_n[l,k]~\equiv~\lim_{\beta\rightarrow+0}\,
   \Delta[\beta^{-1}l,k]\,. \label{3.2}
\eeqa
Furthermore, one can prove the following identity (see Appendix B):
\beqa
   &&\Det^{1/2}\Delta[l+\db l,k+\db k] \nn
   &&~=~\exp\left[-\db\,\int_0^D dt \left\{ 
   \mu_2l(t)\,+\,\frac{1}{24\pi}\,
   \frac{\ldot(t)^2}{l(t)}\right\}\right]\,
   \Det^{1/2}\Delta[l,k]\label{3.5}
\eeqa
with $\mu_2$ some regularization-dependent constant. 
Thus, we have 
\beqa
   &&\Det^{1/2}D_n^\dagger D_n[l+\db l,k+\db k] \nn
   &&=~\lim_{\beta\rightarrow+0}\,
   \exp\left[-\db\,\int_0^D dt \left\{ 
   \frac{\mu_2}{\beta}l(t)\,+\,\frac{1}{24\pi\beta}\,
   \frac{\ldot(t)^2}{l(t)}\right\}\right]\,
   \Det^{1/2}D_n^\dagger D_n[l,k]\,.~~{~}\label{E6}
\eeqa

Combining eqs.\ \eq{E5} and \eq{E6}, and using the invariance of the
measure $\cD_l'k$ under the transformation \eq{E3}, we can evaluate 
the $l$-dependence of the functional $W[\,l\,]$ as follows:  
\beqa
   &&\exp\{-W[\,l+\db l\,]\}\nn
   &&~=~\int\cD_{l+\db l}'k\,\Det^{1/2}D_n^\dagger D_n[l+\db l,k]\nn
   &&~=~\int\cD_{l+\db l}'\left(k+\db k\right)\,
   \Det^{1/2}D_n^\dagger D_n[l+\db l,k+\db k]\nn
   &&~=~\lim_{\beta\rightarrow+0}\,\int\cD_l'k\,
   \exp\left[-\db\,\int_0^D dt \left\{ 
   \frac{\mu_2}{\beta}\,l\,+\,\frac{1}{24\pi\beta}\,
   \frac{\ldot^2}{l}\right\}\right]\,
   \Det^{1/2}D_n^\dagger D_n[l,k]\nn
   &&~=~\lim_{\beta\rightarrow+0}\,
   \exp\left[-\db\,\int_0^D dt \left\{ 
   \frac{\mu_2}{\beta}\,l\,+\,\frac{1}{24\pi\beta}\,
   \frac{\ldot^2}{l}\right\}\right]\,
   \exp\{-W[\,l\,]\}\,.
\eeqa
Namely, 
\beqa
   W[\,l+\db l\,]~=~W[\,l\,]\,+\,\lim_{\beta\rightarrow+0}\,\db\,
   \int_0^D dt \left\{ 
   \frac{\mu_2}{\beta}\,l\,+\,\frac{1}{24\pi\beta}\,
   \frac{\ldot^2}{l}\right\}\,.
\eeqa
Thus, integrating this equation, and using eq.\ \eq{E4}, we finally
obtain 
\beqa
   F[\,l\,]~=~\lim_{\beta\rightarrow+0}\,\exp\,
   \left[\,-\,\int_0^Ddt\,\left\{\mu_2'\,l\,+\,\frac{1}{2\beta}\,
   \frac{\ldot^2}{l}\right\}\right]\,,\label{E7}
\eeqa
where we have replaced $(24\pi\beta)^{-1}\!-\!(32\pi)^{-1}$ by
$(2\beta)^{-1}$ for simplicity, and $\mu_2'$ is a $\beta$-dependent
constant.  

Now we can calculate the cylinder amplitude $f(l',l;D)$ defined in
eq.\ \eq{2.3}.
We first substitute  eqs.\ \eq{F8}, \eq{E1} and \eq{E7} into eq.\
\eq{2.3}, and then obtain
\beqa
   f(l',l;D)~=~\lim_{\beta\rightarrow+0}\,\int \dl\,e^{-\,S_\beta[\,l\,]}
   \,\delta\left(l(t\!=\!0)-l\right)\,\delta\left(l(t\!=\!D)-l'\right)\,,
   \label{3.7}
\eeqa
where
\beqa
   S_\beta[\,l\,]&\equiv&\int dt\,L_\beta(l,\ldot) \nn
   &\equiv&\int dt\,\left( \frac{\ldot^2}{2\beta l}+\mu l\right)\,,
   \label{3.8}
\eeqa
and $\mu\,\equiv\,\mu_0\,+\mu_1\,+\,\mu_2'$ is the renormalized
cosmological constant.

For further calculation it is useful to introduce one-body Hamiltonian
$H$ from this (Euclidian) action $S_\beta[\,l\,]$.
Since the momentum $p$ conjugate to $l$ is
\beqa
   p~\equiv~i\,\frac{\del L_\beta}{\del \ldot}~=~i\,
   \frac{\ldot}{\beta l}\,, \label{3.9}
\eeqa
we can construct $H$ as
\beqa
   H&\equiv&\lim_{\beta\rightarrow+0}\,\left(ip\ldot+L_\beta\right) \nn
   &=&\lim_{\beta\rightarrow+0}\,\left(\frac{\beta}{2} p^2l+\mu l\right) \nn
   &=&\mu l\,.\label{3.10}
\eeqa
Thus, the cylinder amplitude $f(l',l;D)$ is now evaluated to be
\beqa
   f(l',l;D)&=&\const\bra{l'}e^{-DH}\ket{l} \nn
   &=&\const e^{-\mu Dl}\,\delta(l-l')\,. \label{3.11}
\eeqa
The normalization constant is determined to be unity by imposing
the following composition law:
\beqa
   f(l',l;D)~=~\int_0^\infty dl''\,f(l',l'':D_1)\,
   f(l'',l:D-D_1)~~~~(0<D_1<D)\,. \label{3.12}
\eeqa

A comment is now in order.
To make eq.\ \eq{3.12} valid even for the functional integral
expression \eq{3.7}, we have to avoid an overintegration which
possibly occurs at the end points in time ({\it i.e.\,}, boundaries of
cylinder).
However, this may be understood by regarding the functional
measure $\left[dl/l\right]$ as
\beqa
   \dl~\equiv~\prod_{0<t\leq D}\,
   \frac{dl(t)}{l(t)}\,.\label{3.13}
\eeqa
Since we adopt the rule that in eq.\ \eq{3.13} $t=0$ is excluded from
the initial time, the expression $\delta\left(l(t\!=\!0)-l\right)$ 
in eq.\ \eq{3.7} now should be read as 
$\delta\left(l(t\!=\!+0)-l\right)$.

\section{Derivation of $c=0$ string field theory}
\cleqn

In the previous section, we found that in our gauge \eq{2.7} and
\eq{2.8}, the calculation of the cylinder amplitude reduces to that
of quantum mechanics of loop length $l$ described by the Hamiltonian
$H=\mu l$.
In this section, we first evaluate the amplitude for the case in which
a loop splits into two loops.
In fact, such amplitude can be obtained in an almost similar way to
the cylinder one \eq{3.7} except that we have no residual gauge
symmetry at the moment (say, $t=t_0$) when an incoming loop splits
(see Fig.\ 2).
Since such modification is only relevant to the boundary condition of
the tangential component $v^s$ (or zero-modes of the operator 
$D_s^\dagger D_s = - l^{-2}\del_1^2$), the functional measure 
$\cD g_{\mu\nu}/\mbox{Vol(diff)}$ is different from the cylinder one
simply by the factor of the length at $t=t_0$, $l(t_0)$, as is the
case for one-dimensional quantum gravity (see, for example, ref.\
\cite{p2}). 
It should be noted that in the path integration over $l(t)$
(eq.\ \eq{3.13}), we exclude the initial time $t=0$.
Therefore, when one incoming cylinder splits into two outgoing ones,
we multiply the length of the boundary loop of the incoming cylinder.
Similarly, when two incoming cylinders merge into outgoing one,
the lengths of the two boundary loops of the incoming cylinders should
be multiplied (see Fig.\ 3).

In summary, we have the following Feynman rule for our loop
dynamics:\\

{}~~\\
\underline{propagator\,}:\\

\begin{picture}(100,100)(-150,0)
   \put(0,30){\line(0,1){40}}
   \put(20,50){\makebox(0,0)[bl]{$=~\mu\,l\,\delta(l-l')$}}
   \put(0,20){\makebox(0,0){$l$}}
   \put(0,80){\makebox(0,0){$l'$}}
\end{picture}

{}~~\\
\underline{loop-splitting vertex\,}:\footnote{
The following three figures should be understood to represent
graphs with external lines (cylinders) amputated.}\\

\begin{picture}(100,80)(-150,0)
   \put(0,50){\line(0,-1){20}}
   \put(0,50){\line(4,3){20}}
   \put(0,50){\line(-4,3){20}}
   \put(20,56){\makebox(0,0){$l'$}}
   \put(-20,56){\makebox(0,0){$l$}}
   \put(10,40){\makebox(0,0){$l''$}}
   \put(40,50){\makebox(0,0)[bl]{$=~l''\,\delta(l+l'-l'')$}}
\end{picture}

{}~~\\
\underline{loop-merging vertex\,}:\\

\begin{picture}(100,100)(-150,-20)
   \put(0,50){\line(0,1){20}}
   \put(0,50){\line(4,-3){20}}
   \put(0,50){\line(-4,-3){20}}
   \put(20,44){\makebox(0,0){$l'$}}
   \put(-20,44){\makebox(0,0){$l$}}
   \put(10,60){\makebox(0,0){$l''$}}
   \put(40,50){\makebox(0,0)[bl]{$=~l\,l'\,\delta(l+l'-l'')$}}
\end{picture}

To go into the second quantization formalism,
we further need to introduce the cap amplitude $\rho(l)$ (amplitude
for instantaneous vanishing of the loop with length $l$)\,:\\

{}~~\\
\underline{cap\,}:\\

\begin{picture}(0,80)(-150,0)
   \put(0,35){\line(0,1){30}}
   \put(30,50){\makebox(0,0)[bl]{$=~\rho(l)$}}
   \put(8,50){\makebox(0,0){$l$}}
   \put(0,70){\makebox(0,0){$\bigotimes$}}
\end{picture}

Using the above rule we now can write down the second-quantized
Hamiltonian in the following form:
\beqa
   \cH&=&\mu\,\int dl\,l\,\Psid(l)\,\Psi(l) \nn
   &&~~~+\,\int dl\,dl'\,(l+l')\,\Psid(l)\,\Psid(l')\,\Psi(l+l') \nn
   &&~~~+\,g\,\int dl\,dl'\,l\,l'\,\Psid(l+l')\,\Psi(l)\,\Psi(l') \\
   &&~~~+\,\int dl\,\rho(l)\,\Psi(l)\,. \n
\eeqa
Here we have set to unity the coefficients of the second and forth
terms by appropriately rescaling the Hamiltonian $\cH$ (or
equivalently, its conjugate geodesic-time) and the operators
$\Psi(l)$, $\Psid(l)$. 
Furthermore, we can eliminate the first term (kinetic term) by making
a shift of $\Psid(l)$ as
\beqa
   \Psid(l)~\mapsto~\Psid(l)\,-\,\frac{\mu}{2}\,\delta(l)
\eeqa
and formally using the identity $l\,\delta(l)\,=\,0$.
Finally the Hamiltonian reads
\beqa
   \cH&=&\int dl\,dl'\,(l+l')\,\Psid(l)\,\Psid(l')\,\Psi(l+l') \nn
   &&~~~+\,g\,\int dl\,dl'\,l\,l'\,\Psid(l+l')\,\Psi(l)\,\Psi(l') \\
   &&~~~+\,\int dl\,\rho(l)\,\Psi(l)\,. \n
\eeqa

If we could prove that $\rho(l)\,=\,\delta''(l)-\mu \delta(l)$, then
we would succeed in reproducing the Hamiltonian for $c=0$ string field
theory of ref.\ \cite{ik}, and so all the results of $c=0$ matrix model.
However, at this stage we have no definite proof for this
statement.\footnote{
We can still make a handwaving argument as follows:
First, we consider a cylinder with two loop boundaries of length $l$
and $0$.
Due to eq.\ \eq{3.11}, the amplitude $f(0,l;D)$ for such configuration
has its support only at $l=0$.
On the other hand, the cylinder can be obtained from a tiny (almost
flat) disk by making a hole on it.
However, there exists arbitrariness for the location of the hole, and
such arbitrariness should be proportional to the area of the disk.
Thus, it is plausible to assume that $f(0,l;D)\,\sim\,D l\,\rho(l)$
for small $D$ and $l$, or
\beqa
   \rho(l)&\sim&\frac{1}{Dl}\,f(0,l;D) \nn
   &=&\frac{1}{Dl}\,e^{-\mu D l}\,\delta(l)\,. \n
\eeqa
Furthermore, $D$ should scale as $l$ in the limit $l\,\rightarrow\,0$,
and thus we obtain
\beqa
   \rho(l)&\sim&\frac{1}{l^2}\,e^{-\mu l^2}\,\delta(l) \nn
   &\sim&\delta''(l)\,-\,\mu\,\delta(l)\,. \n
\eeqa
}

\section{Conclusion}
\cleqn

In this paper, we have shown that pure gravity reduces to
quantum mechanics of loop length $l$.
Furthermore, we have demonstrated the vanishing of the kinetic term
in the second-quantized form, which yields the pregeometric
Hamiltonian for $c=0$ string field theory.

Besides the elaboration of our present formulation, one of the
intriguing problems is to extend our system to the ones in which
gravity is coupled to matters.
In particular, we should identify physical operators and investigate
their structures.
An intuitive conjecture at this stage is that loops will be further
labeled with conformal primary fields, and that all the physical
operators known so far are obtained by expanding such loops in their
loop length.
Investigation along this line is now in progress, and will be
reported elsewhere.

\section*{Acknowledgements}
\cleqn

We are grateful to E.\ D'Hoker, T.\ Kawano, R.\ Nakayama, Y.\
Okamoto and K.\ Yoshida for useful discussions.
This work is supported in part by the Grant-in-Aid for Scientific
Research from the Ministry of Education, Science and Culture.

\setcounter{section}{0}
\renewcommand{\thesection}{Appendix \Alph{section}}

\section{Proof of eq.\ (2.19)}
\renewcommand{\theequation}{A.\arabic{equation}}
\cleqn


In general, the norm \eq{2.4} of infinitesimal deformation of metric
can be rewritten into the following form if we use the ADM
decomposition \eq{2.2}:
\beqa
   \norm{\delta g_{\mu\nu}}_g~=~4\,\int d^2x\,N\sqrt{h}\,\left[\,
   \left(\frac{\delta h}{2h}\right)^2\,+\,
   \left(\frac{\delta N}{N}\right)^2\,+\,
   \frac{1}{2}\,\left(\frac{\sqrt{h}}{N}\,\delta\lambda\right)^2
   \,\right]\,.
   \label{A.1}
\eeqa
In particular, around $\gbar_{\mu\nu}$ it is
expressed as
\beqa
   \norm{\delta g_{\mu\nu}}_\gbar~=~\int d^2x\,l\,\left[\,
   \left(\frac{\delta h}{2l^2}\right)^2\,+\,
   \left(\delta N\right)^2\,+\,
   \frac{1}{2}\,\left(l\,\delta\lambda\right)^2\,\right]\,,
   \label{A.2}
\eeqa
where we have omitted the irrelevant coefficient ``4'' from the expression.
Each of the terms appearing in eq.\ \eq{A.2} can be calculated by
using eq.\ \eq{2.15} and found to be
\beqa
   \frac{\delta h}{2l^2}&=&
   \frac{\delta l}{l}\,-\,\omega\,\delta v^n\,+\,D_s\,\delta v^s\,, \nn
   \delta N &=& D_n \,\delta v^n \,, \label{A.3} \\
   l\,\delta\lambda&=&
   l\,\delta k\,+\,D_s\,\delta v^n\,+\,\left(D_n+\omega\right)\,\delta
   v^s \n 
\eeqa
with $\omega=k'-\ldot/l$.
Thus, we obtain
\beqa
   \norm{\delta g_{\mu\nu}}_\gbar&=&\int d^2x\,l\,\left[\,
   \left(\frac{\delta l}{l}-\omega\,\delta v^n+D_s\,\delta
   v^s\right)^2\,+\,\left(D_n\,\delta v^n\right)^2\right. \nn
   &&\left.~~~~~~+\,\frac{1}{2}\,\left(l\,\delta k+D_s\,\delta v^n
   +(D_n+\omega)\,\delta v^s\right)^2\,\right]\,. \label{A.4}
\eeqa
The above expression can be transformed into the form of eq.\
\eq{2.16} by making a shift of $\delta v^n$ and $\delta v^s$.
In doing so, there might occur some problems concerned with
zero-modes of the operators $D_n$ and $D_s$.
However, the zero-modes (constant modes) of $D_n$ always vanish since
we have used the Dirichlet boundary condition for $D_n$, while the
zero-modes of $D_s$ were extracted in the beginning in order
to further fix the residual gauge symmetry (see the discussion
following eq.\ \eq{2.20}). \\

\section{Proof of eq.\ (3.12)}
\renewcommand{\theequation}{B.\arabic{equation}}
\cleqn

First, we define the determinant of the Laplacian
\beqa
   \Delta\,[g_{\mu\nu}]~\equiv~-\frac{1}{\sqrt{g}}
   \del_\mu(\sqrt{g}g^{\mu\nu}\del_\nu) \label{B.1}
\eeqa
by the following equation:
\beqa
   \ln \Det \Delta\,[g_{\mu\nu}]~\equiv~-\,\int_{\epsilon^2}^\infty
   \frac{ds}{s}\,\Tr\,e^{-s\Delta\,[g_{\mu\nu}]} \label{B.2}
\eeqa
with ultra-violet cutoff $\Lambda\sim\epsilon^{-1}$.
Then the determinant will change under the infinitesimal Weyl
transformation $g_{\mu\nu}\mapsto e^{2\db\sigma}g_{\mu\nu}$
as\footnote{
Here for the scalar curvature $R$ we used the following convention:
\beqa
   \int d^2x\,\sqrt{g}\,R~=~4\pi\,\chi~=~8\pi\,(1-h)\,, \n
\eeqa
which is different from that in ref.\ \cite{dp} by 2.
}
\beqa
   \db\,\ln\Det\Delta\,[g_{\mu\nu}]
   &\equiv&\ln\Det\Delta\,[e^{2\db\sigma}g_{\mu\nu}]\,-\,
   \ln\Det\Delta\,[g_{\mu\nu}] \nn
   &=&-\,2\int d^2x\,\sqrt{g}\,\db\sigma\,\left(
   \frac{1}{4\pi\epsilon^2}\,+\,\frac{1}{24\pi}R\,[g_{\mu\nu}]\,+\,
   O(\epsilon^2)\right)\,. \label{B.3}
\eeqa

Our strategy to prove eq.\ \eq{3.5} is to relate the infinitesimal
deformation $\db l(t)$ of loop length at time $t$, to some
infinitesimal Weyl transformation $\db\sigma(t)$ which depends only
on $t$.
However, since Weyl transformations generally break the gauge
condition \eq{2.7}, a proper reparametrization $x^\mu\mapsto
\tilde{x}^\mu(x)$ has to be accompanied,
\beqa
   ds^2&=&e^{2\db\sigma(t)}\gbar_{\mu\nu}dx^\mu dx^\nu \nn
   &=&e^{2\db\sigma(t)}\,\left[
   (dt)^2\,+\,l(t)^2\left(k(t,x)dt+dx\right)^2\right] \label{B.4}\\
   &\equiv&(d\tilde{t})^2\,+\,\tilde{l}(\tilde{t})^2\left(
   \tilde{k}(\tilde{t},\tilde{x})d\tilde{t}+d\tilde{x}\right)^2\,.\n
\eeqa
One can here easily show that the following reparametrization
satisfies the above requirement:
\beqa
   \tilde{t}(t,x)&\equiv&\int_0^tdt'\,e^{\db\sigma(t')}\,, \nn
   \tilde{x}(t,x)&\equiv&x\,,\label{B.5}
\eeqa
with
\beqa
   \tilde{k}(\tilde{t},\tilde{x})&=&e^{-\db\sigma(t)}k(t,x)\,, \nn
   \tilde{l}(\tilde{t})&=&e^{\db\sigma(t)}l(t)\,.\label{B.6}
\eeqa
Note that if we introduce the symbol
\beqa
   \db\rho(t)&\equiv&\tilde{t}\,-\,t\nn
   &=&\int_0^tdt'\,\db\sigma(t')\,,\label{B.7}
\eeqa
then the following equation holds:
\beqa
   \db\rho(0)~=~\db\rho(D)~=~0\,,\label{B.8}
\eeqa
since two boundaries of the cylinder in question are
parametrized as $t=\tilde{t}=0$ and $t=\tilde{t}=D$.

By using eqs.\ \eq{B.6} and \eq{B.7}, $\db k(t,x)$ and $\db l(t)$ can
now be expressed by $\db\rho(t)$ as follows:
\beqa
   \db k(t,x)&\equiv&\tilde{k}(t,x)\,-\,k(t,x) \nn
   &=&-\,\dot{k}(t,x)\,\db\rho(t)\,-\,k(t,x)\,\db\dot{\rho}(t) \nn
   &=&-\,\frac{\del}{\del t}\left(k(t,x)\,\db\rho(t)\right)\,, \\
   \db l(t)&\equiv&\tilde{l}(t)\,-\,l(t) \nn
   &=&-\,\ldot(t)\,\db\rho(t)\,+\,l(t)\,\db\dot{\rho}(t) \nn
   &=&l(t)^2\,\frac{d}{dt}\left(\frac{\db\rho(t)}{l(t)}\right)\,.
   \label{B.9}
\eeqa
These are nothing but the transformation introduced in sect.\ 3. 
Conversely, $\db\rho(t)$ can be expressed by $\db l(t)$ as
\beqa
   \db\rho(t)~=~l(t)\,\int_0^tdt'\,\frac{\db l(t')}{l(t')^2}\,,
   \label{B.10}
\eeqa
and we obtain
\beqa
   \db\sigma(t)&=&\db\dot{\rho}(t) \nn
   &=&\ldot(t)\int_0^tdt'\,\frac{\db l(t')}{l(t')^2}\,+\,
   \frac{\db l(t)}{l(t)}\,. \label{B.11}
\eeqa
Note that eq.\ \eq{B.8} implies that
\beqa
   \int_0^Ddt\,\frac{\db l(t)}{l(t)^2}~=~0\,.\label{B.12}
\eeqa
On the other hand, in our temporal gauge the scalar curvature $R$ is
given by
\beqa
   R\,[\gbar_{\mu\nu}]~=~-\,2\,\frac{\ddot{l}}{\,l\,}\,+\,
   2\left(\,\dot{k}-kk'+2k\,\frac{\ldot}{\,l\,}\,\right)^{\prime}\,.
   \label{B.13}
\eeqa
Thus, substituting eqs.\ \eq{B.11} and \eq{B.13} into eq.\ \eq{B.3},
we obtain
\beqa
   &&\db\,\ln\Det\Delta\,[\gbar_{\mu\nu}] \nn
   &&~~=~\int_0^Ddt\int_0^1dx \,\left[-\,2\mu_2\,l(t)
   \begin{array}{c}
   ~ \\ ~
   \end{array} \right.\nn
   &&~~~~~~~~~~+\,\frac{1}{6\pi}\left.\left\{
   \ddot{l}(t)\left(\ldot(t)\int_0^tdt'\,\frac{\db l(t')}{l(t')^2}
   \,+\,\frac{\db l(t)}{l(t)}\right)\,+\,
   \mbox{(total derivative in $x$)}\right\}\right] \nn
   &&~~=~\int_0^Ddt \left[ -\,2\mu_2\,l(t)\,+\,\frac{1}{6\pi}
   \ddot{l}(t)\left(\ldot(t)\int_0^tdt'\,\frac{\db l(t')}{l(t')^2}
   \,+\,\frac{\db l(t)}{l(t)}\right)\right]\,,\label{B.14}
\eeqa
where $\mu_2\equiv1/8\pi\epsilon^2$. 
This expression can be further simplified by using eq.\ \eq{B.12},
and becomes
\beqa
   \db\,\ln\Det\Delta\,[\gbar_{\mu\nu}]~=~-\,
   \db\int_0^Ddt\,\left[\,2\mu_2\,l(t)\,+\,
   \frac{1}{12\pi}\,\frac{\ldot(t)^2}{l(t)}\,\right]\,.
   \label{B.15}
\eeqa
After integrating this equation, we finally obtain eq.\ \eq{3.5}.

\newpage

\section*{Figure captions}
\cleqn

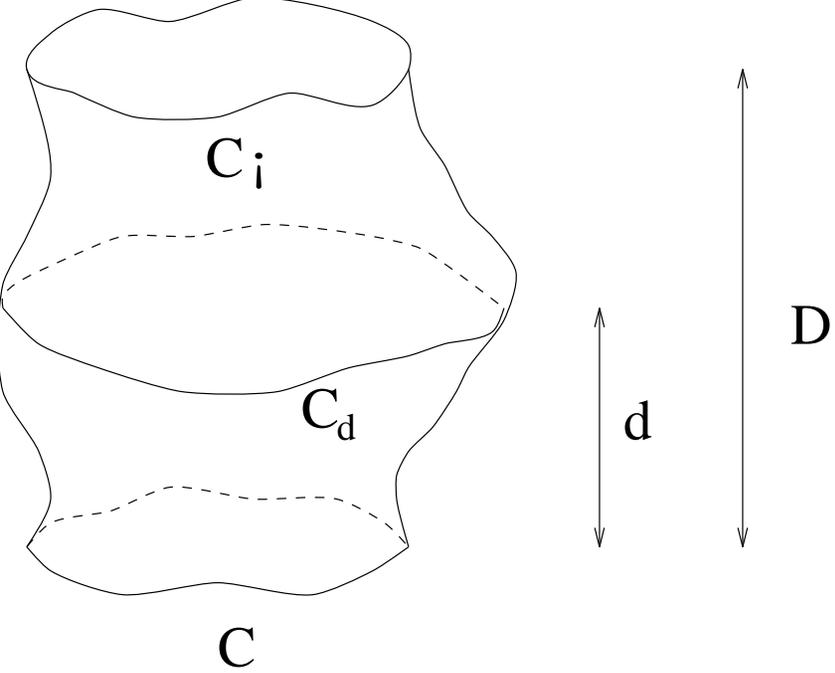
\noindent{\bf Fig.\ 1}\\

A cylinder $M$ with two loop boundaries $C$ and $C'$.
$C_d$ is the subset in $M$ consisting of the points whose geodesic
distance from $C$ is $d$.
We here assume that $C_d$ is homeomorphic to $C$ for any
$d\,\in\,[0,D]$ and $C_D\,=\,C'$.  \\
{}~~\\

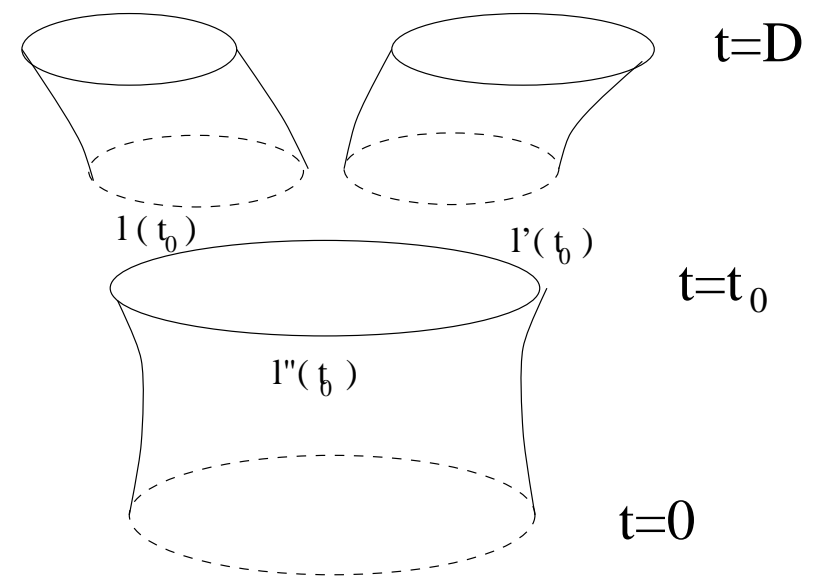
\noindent{\bf Fig.\ 2}\\

A loop splits into two loops at $t=t_0$.
Here $f(t_0)\,\equiv\,l''(t_0)\,\delta(l(t_0)+l'(t_0)-l''(t_0))$. \\
{}~~\\

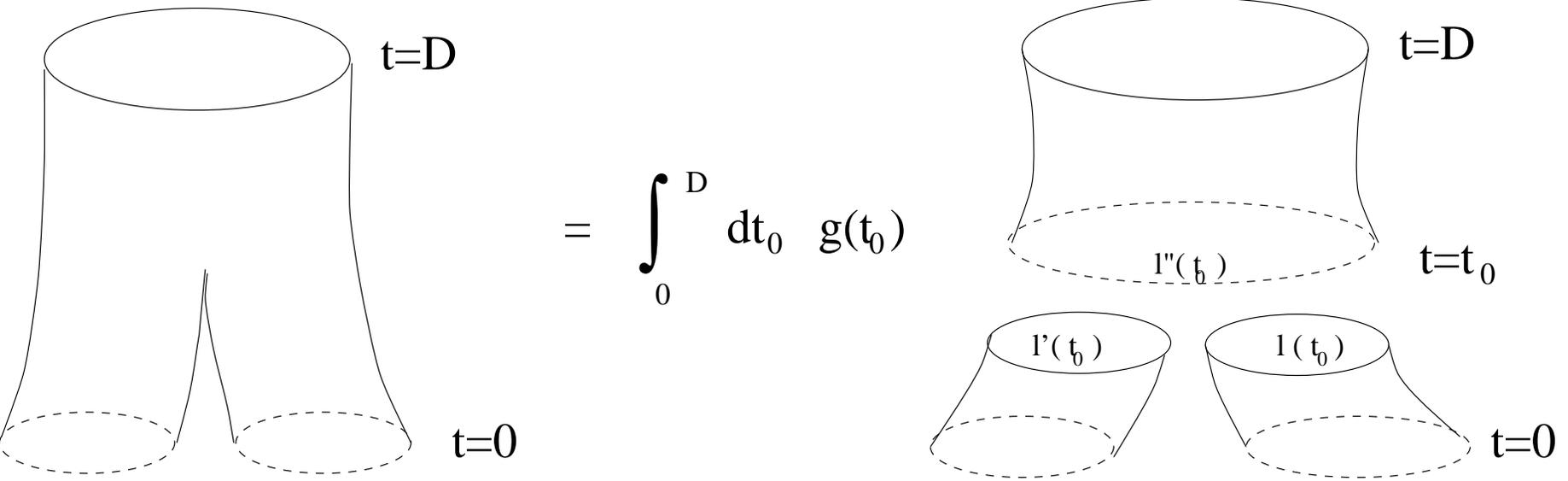
\noindent{\bf Fig.\ 3}\\

Two loops merge into a single loop at $t=t_0$.
Here $g(t_0)\,\equiv\,l(t_0)\,l'(t_0)\,\delta(l(t_0)+l'(t_0)-l''(t_0))$.



\end{document}